\documentstyle[anta, twoside]{article}

\input{psfig.sty}

\def\Journal#1#2#3#4{(#1) {#2} {\bf #3}, #4}

\def\AAp{{\em Astron. Astrophys.}}
\def\AJ{{\em Astron.~J.}}
\def\ApJ{{\em Astrophys.~J.}}

\def\ApJSS{\em Astrophys.~J. Suppl.}
\def\AAps{\em Astron. Astrophys. Suppl.}

\newcommand{\HII}{{\rm H\,\scriptstyle II}}
\newcommand{\lamA}{{\rm $\lambda\,21\,$cm}}
\newcommand{\degs}{{$^{\circ}$}}

\renewcommand{\thefootnote}{\fnsymbol{footnote}}
\begin{document}

\markboth{M. Wolleben \& W. Reich}{Modelling Faraday Screens}

\thispagestyle{plain}
\setcounter{page}{99}

\title{Modelling Faraday Screens in the\\Interstellar Medium\footnote[1]{Based on
observations with the Effelsberg 100-m telescope operated by the
Max-Planck-Institut f\"ur Radioastronomie (MPIfR), Bonn, Germany} }

\author{Maik Wolleben and Wolfgang Reich}

\address{Max-Planck-Institut f\"ur Radioastronomie,
Auf dem H\"ugel 69, 53121 Bonn, Germany}


\maketitle

\abstract{ Maps of Galactic polarized continuum emission at $1408$,
$1660$, and $1713$~MHz towards the local Taurus molecular cloud complex
were made with the Effelsberg 100-m telescope. Minima in the polarized
emission which are located at the boundary of a molecular cloud were
detected. Beside high rotation measures and unusual spectral indices of
the polarized intensity, these features are associated with the molecular
gas. At the higher frequencies the minima get less distinct. We have
modelled the multi-frequency observations by placing magneto-ionic
Faraday screens at the distance of the molecular cloud. In this model
Faraday rotated background emission adds to foreground emission
towards these screens. The systematic variation of the observed
properties is the result of different line-of-sight lengths through
the screen assuming spherical symmetry. For a distance of $140$~pc to
the Taurus clouds the physical sizes of the Faraday screens are of the
order of $2$~pc. In this paper we describe the data calibration and
modelling process for one such object. We find an intrinsic rotation
measure of about $-29$~rad$\,$m$^{-2}$ to model the observations. It
is pointed out that the observed rotation measure differs from the
physical. Further observational constraints from H$\alpha$ observations
limit the thermal electron density to less than $0.8$~cm$^{-3}$, and we
conclude that the regular magnetic field strength parallel to the
line-of-sight exceeds $20~\mu$G to account for the intrinsic rotation
measure.
}

\renewcommand{\thefootnote}{\arabic{footnote}}
\section{Introduction}

Various surveys of Galactic polarized emission revealed an unexpected
richness in highly varying structures in the polarized sky as discussed
on this conference. In many cases these fluctuations in the polarized
intensity and position angle have no counterpart in total intensity.
One likely explanation for these observations is polarized background
emission modulated by Faraday rotation. However, many emitting and
rotating layers may exist along the line-of-sight so that the
observed polarization is a superposition of modified background and
unmodified foreground emission layers.

The average density of thermal electrons in the Galactic plane is
about $0.03$~cm$^{-3}$ and the average of the regular magnetic field
along the line-of-sight is about $1$ to $2~\mu$G (Taylor~\&~Cordes
1993, Gom\'ez et al. 2001). However, local enhancements of
$n_{\mathrm{e}}$ are often observed as diffuse $\HII$-regions due to
their optical H$\alpha$ emission occurring from ionization and
recombination of hydrogen. At low observing frequencies Faraday
rotation is high and the polarization angle of synchrotron radiation
is very sensitive to fluctuations in the electron density, Other than
H$\alpha$ emission Faraday rotation depends on electrons from all
sorts of elements. Measurements of local conditions of the Galactic
magnetic field are not straightforward and often done by exploiting
the Zeeman splitting effect. Faraday rotation of polarized radiation
is another tool for the investigation of magnetic fields in case the
thermal electron density is known.

For a given observing frequency the amount of Faraday rotation is
proportional to the product of the electron density times the magnetic
field component parallel to the line-of-sight. Fluctuations in either
or both lead to changes in the observed polarization angle. Since such
fluctuations are often of small spatial extent, one can describe them in
terms of Faraday screens. However, the observation of the effect of
Faraday screens on polarized background radiation is not
straightforward since foreground emission adds vectorially to the
modified background. The closer the screen the more apparent its effect,
and the key problem is the unknown distance of the Faraday screens.
However, distances to the Taurus--Auriga molecular cloud complexes are
known to be about $140 \pm 20$~pc (e.g. Elias 1978) and structures on
pc--scales can be resolved with arcmin angular resolution. In addition
the Taurus--Auriga complex is located at medium latitudes well below the
Galactic plane resulting in a relatively short line-of-sight through the
Galaxy.

Here, we analyze a map from the \lamA~{\it Effelsberg Medium Latitude
Survey} (EMLS, see Reich et al., this volume), which shows a number of
enhancements and depressions in the polarized intensity  apparently
related to molecular gas of the Taurus complex. We interpret the
coincidence in position and shape as strong evidence for Faraday
effects taking place at the distance of the molecular cloud. In order
to derive physical properties of the associated Faraday screens we
have complemented the\lamA~survey data of the Taurus--Auriga region by
observations at $18$~cm wavelength. We modelled the polarization
data in order to constrain the physical parameters of the Faraday
screens by taking foreground and background emission into account.

\section{Observation and Calibration}

The \lamA~EMLS covers the northern Galactic plane in the range of $|b|
\le 20$\degs~at a frequency of $1.4$~GHz. Follow--up observations of a
field north of the center of the Taurus molecular cloud complex were
carried out at $1660$ and $1713$~MHz. Total intensities and linear
polarization were measured simultaneously with sensitivities of $15$~mK
($1408$~MHz) to $19$~mK ($1713$~MHz) for Stokes $I$ and $8$~mK
($1408$~MHz) to $10$~mK ($1713$~MHz) for Stokes $U$ and $Q$ at angular
resolutions of $9\farcm 35$~($1408$~MHz) to
$7\farcm 87$~($1713$~MHz). The same receiver was used for all
three frequencies, but different HF-filters were selected to suppress
interferences. The effective bandwidth was $20$~MHz for $1408$~MHz and
$14$~MHz for $1660$ and $1713$~MHz. Fields were mapped two times in
orthogonal scanning directions and are fully sampled. $3$C$\,286$
served as the main calibrator for total power and polarization data.

Varying ground and atmospheric radiation causes temperature gradients
across the map. In order to remove such gradients a linear baseline is
subtracted from each subscan. This procedure also removes real sky
signals of large extent, which leads to a similar problem like missing
zero spacings in synthesis telescope imaging. At $1408$~MHz the missing
large-scale emission is usually recovered by absolutely calibrated
data. For total intensities 1.4~GHz data from the Stockert northern-sky
survey (Reich 1982, Reich \& Reich 1986) were used. For polarization
data we rely so far on the Dwingeloo survey (Brouw \& Spoelstra 1976).
The final calibration of polarization will be made with the data from
the new DRAO 1.4~GHz survey (see Wolleben et al., this volume).
Therefore all quantitative results given below should be taken as
preliminary. However, other than at 1.4~GHz there exist no absolutely
calibrated surveys at $1660$ or $1713$~MHz and we calibrated the maps
in the following way: The temperature spectral index $\beta$
($T\propto\nu^{\beta}$) of Galactic continuum emission was adopted to
be $\beta=-2.7$ in the Taurus area (Reich \& Reich 1988). We assumed
the same spectral index also for the large-scale polarized intensity
and calculated an average offset for the $1660$ and $1713$~MHz maps
from the $1408$~MHz map. Rotation measures across the Taurus region
were determined by Spoelstra (1972) and are very low everywhere
varying around zero, and therefore we assumed no position angle
difference for the large-scale emission.

\begin{figure}[tb]
\centerline{\psfig{figure=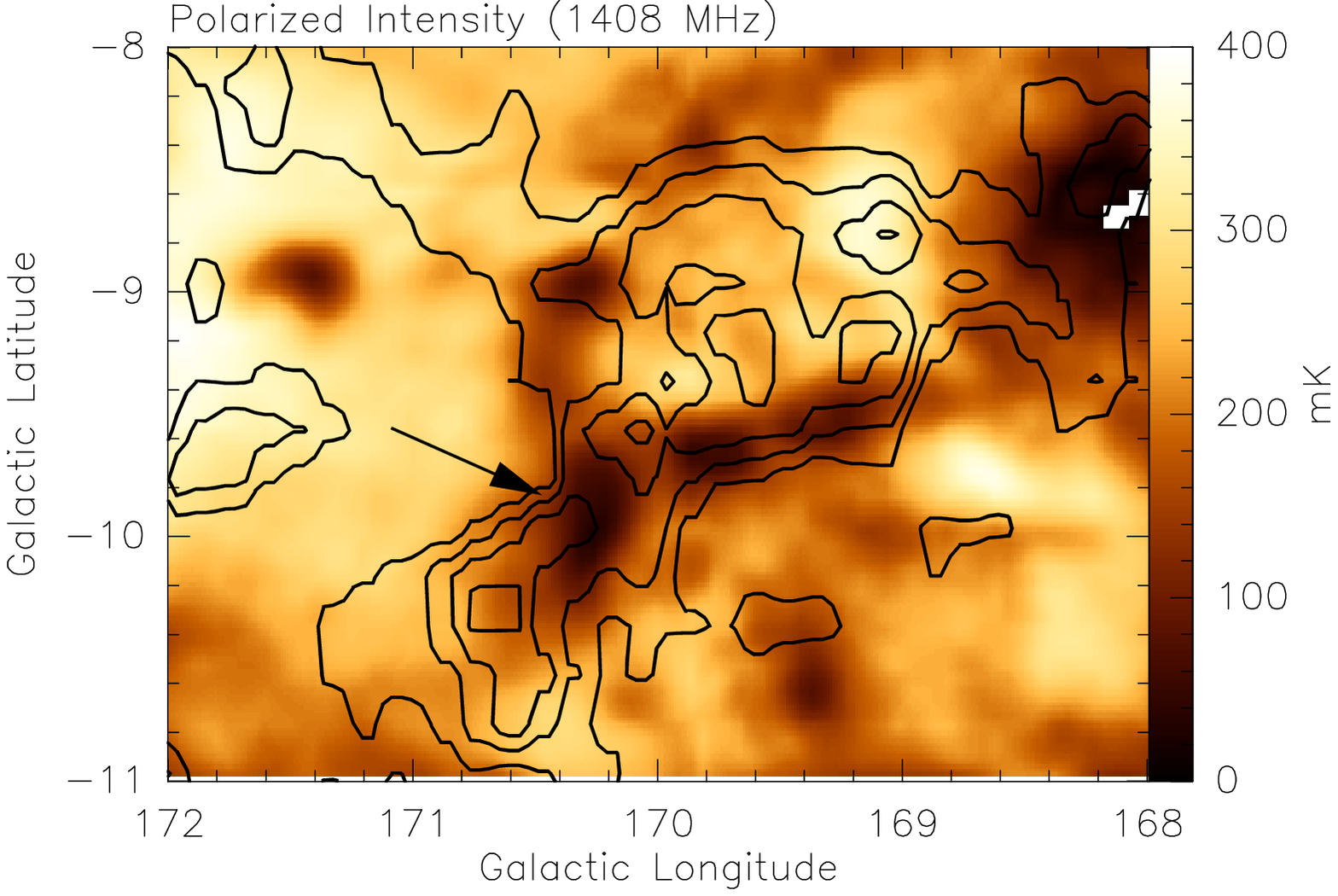,angle=270,clip=T,width=10.2cm,bbllx=0pt,bblly=330pt,bburx=302pt,bbury=783pt}}
\centerline{\psfig{figure=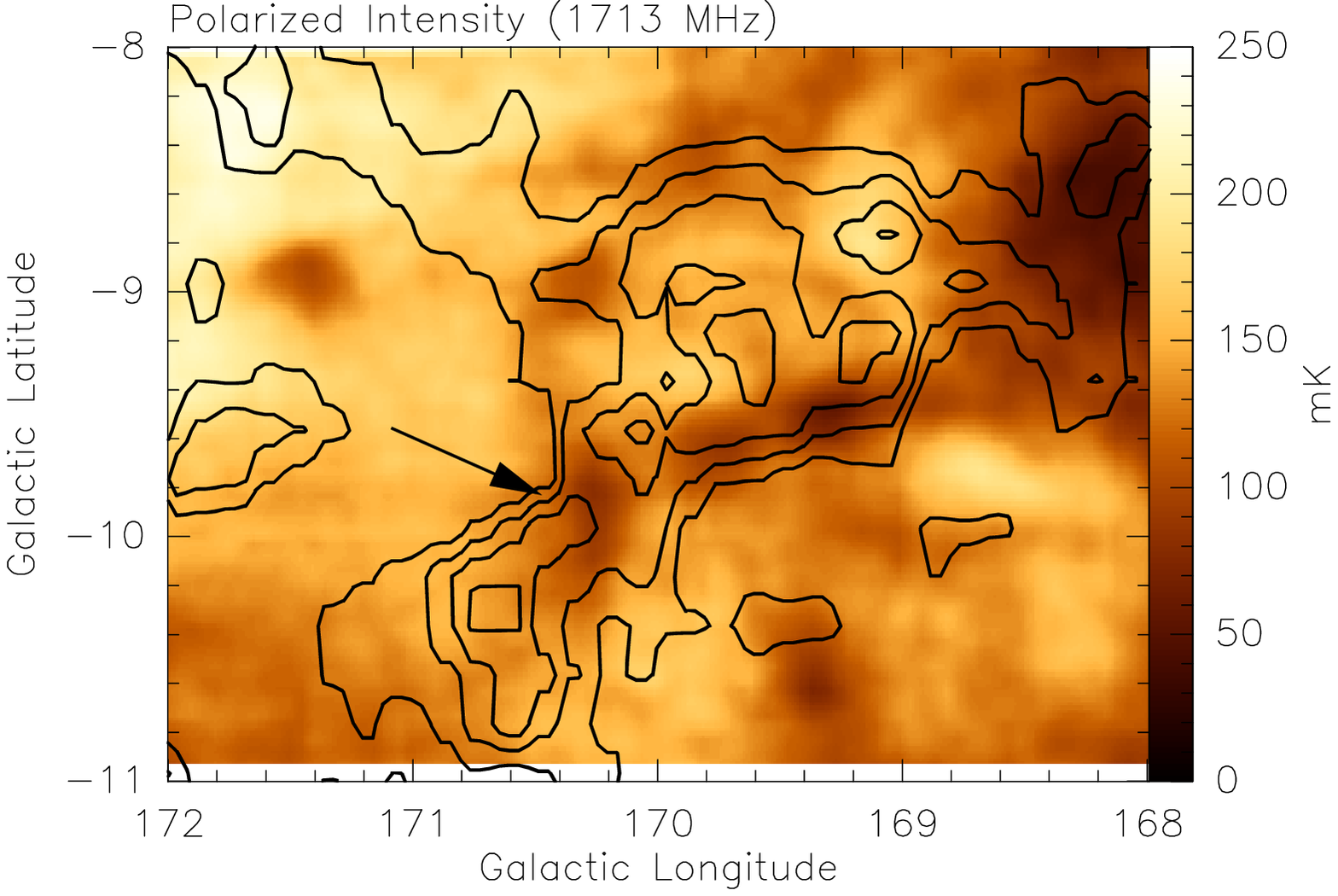,angle=270,clip=T,width=10.2cm,bbllx=-15pt,bblly=330pt,bburx=302pt,bbury=783pt}}
\caption{Maps of the polarized intensity at $1408$ (top) and
$1713$~MHz (bottom) towards a $4\times3$\degs~region north of the
center of the Taurus molecular cloud complex. Contours indicate the
intensity of the velocity-integrated brightness temperature of
$^{12}$CO(1-0) emission (Dame et al. 2001). Contour levels are from 9
to 50~K$\,$km$\,$s$^{-1}$ in steps of 3~K$\,$km$\,$s$^{-1}$. The
minimum in the polarized intensity modelled as a Faraday screen in
Sect. 3 is marked.}
\label{mapchen}
\end{figure}

\begin{figure}[tb]
\centerline{\psfig{figure=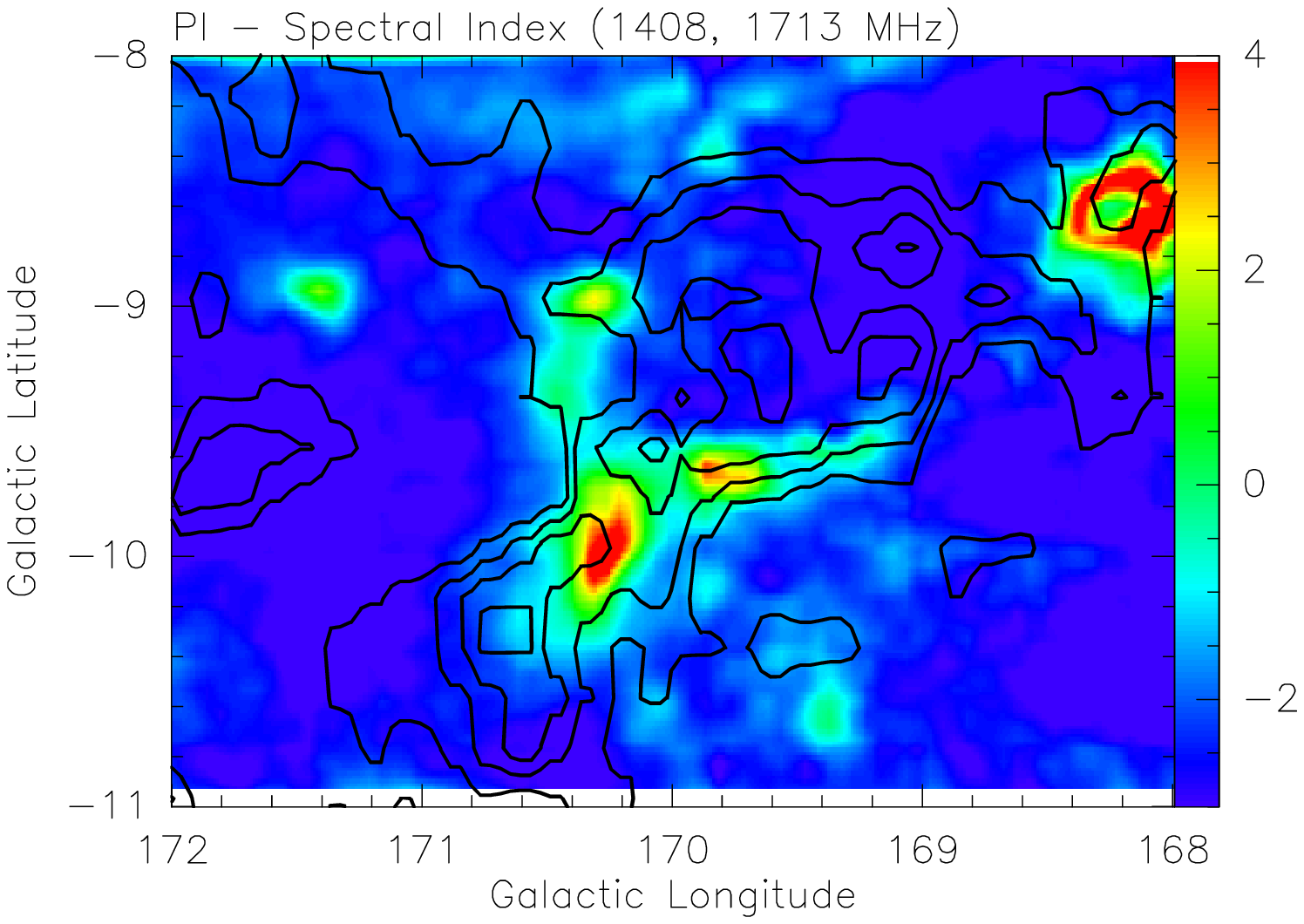,angle=270,clip=T,width=10.2cm,bbllx=72pt,bblly=13pt,bburx=388pt,bbury=464pt}}
\centerline{\psfig{figure=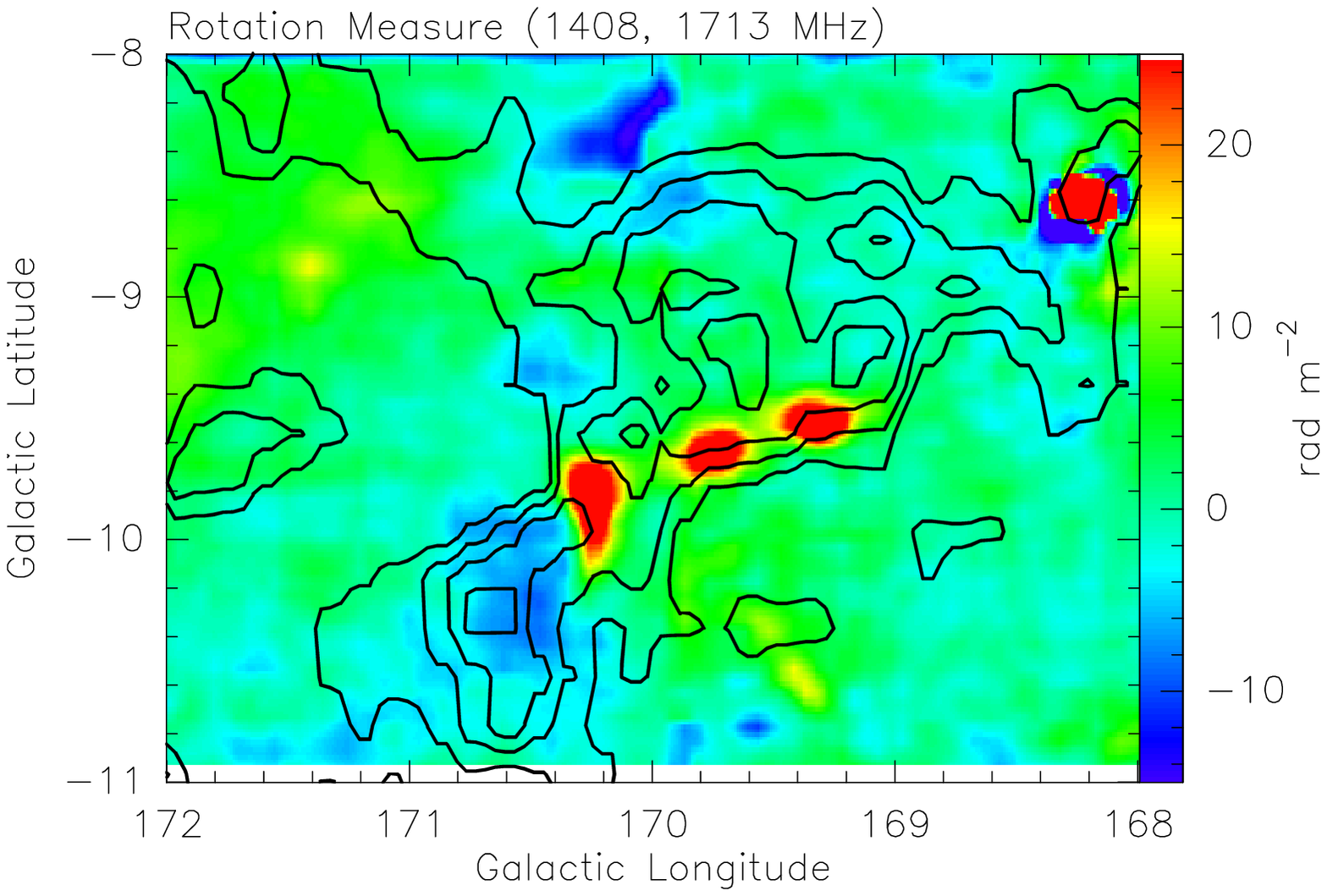,angle=270,clip=T,width=10.2cm,bbllx=72pt,bblly=13pt,bburx=374pt,bbury=464pt}}
\caption{Maps show the spectral index distribution of the polarized
intensity (top) and the rotation measure distribution derived from the
polarization angle rotation between $1408$ and $1713$~MHz (bottom).
Contours are the same as in Fig.~\ref{mapchen}.}
\label{mapchen2}
\end{figure}

The total power maps for all three frequencies show smooth diffuse
emission varying mainly with Galactic latitude and a large number of
unrelated extragalactic sources, but no structures related to the
polarized emission. However, there are numerous small-scale
polarization minima obviously related to the molecular gas
cloud (see Fig.~\ref{mapchen}). They show rather clear differences in
intensity and polarization angle distribution between $1408$ and
$1713$~MHz. Polarized intensities towards these minima increase at
higher frequencies, which is in contrast to the large-scale polarized
emission and other obviously unrelated small-scale variations. The
variation of the polarization angle with frequency reveals rotations
measures of up to $50$~rad$\,$m$^{-2}$ (see Fig.~\ref{mapchen2}). The
$1660$~MHz data were used mainly for a consistency check of the
modelling as described below.

\section{Modelling a Faraday Screen}

We have applied a simple model to describe the observed variations in
the polarization maps at all three frequencies. In this model, a
Faraday screen is modulating background polarized emission passing
through it, which adds to a constant foreground emission. The
foreground and background emission is assumed to be constant for all
lines-of-sight through the Faraday screen. This seems justified since
variations of polarized emission outside the Faraday screens are on
larger scales.

\begin{figure}[tb]
\centerline{\psfig{figure=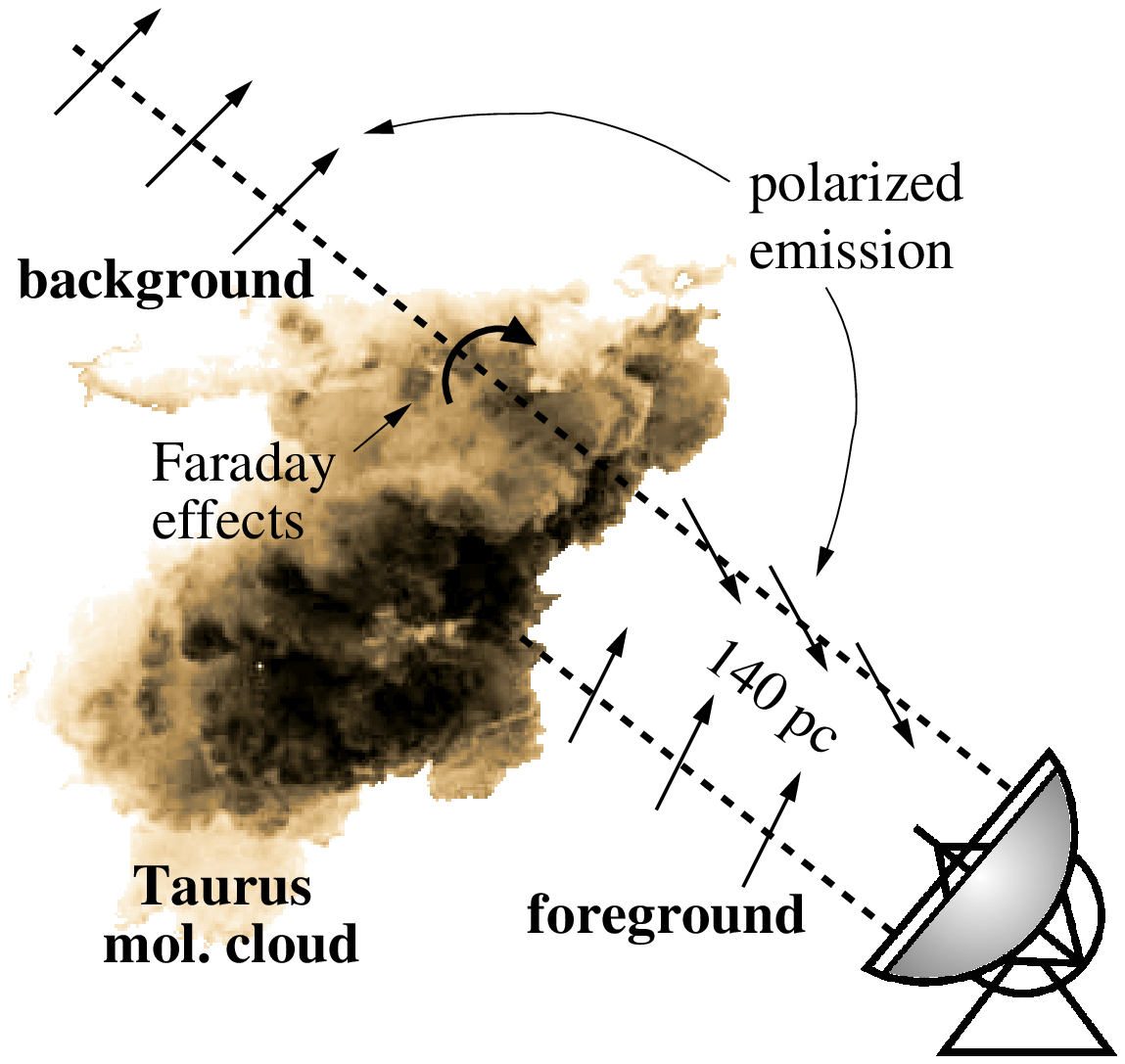,width=10cm}}
\caption{Sketch of a Faraday screen showing the configuration of
foreground and background polarized emission. Here, the Faraday screen
is located at the surface of a molecular cloud. The observer measures
the superposition of polarized emission from an unmodified foreground
and a Faraday rotated and possibly depolarized background. The
distance to the Taurus molecular clouds is $140$~pc. }
\end{figure}

At first we define a Faraday screen as an object which can affect the
position angle and polarized intensity of polarized background
radiation. The effect of the screen will depend on the path length
radiation has to pass through it and thus depends on its size and shape.
For reasons of simplicity we assume Faraday screens to be spherical
objects with constant electron density, homogeneous magnetic field, and
radius $R$. In case of elliptically shaped objects, coordinates were
transformed to allow modelling in circular coordinates. The path
length through the screen will therefore vary systematically with the
observed position. With $r$ as the distance from the center of the
screen projected on the sky, the path length $L$ is then given by
$L(r)=2R\cdot(1-\frac{r^2}{R^2})^{1/2}$ and the fractional path length
$l(r)=\frac{L(r)}{2R}$.

A Faraday screen may decrease the degree of polarization by beam
depolarization, which seems possible due to the relatively small
spatial extent of the Faraday screens discussed here. We assume any
depolarization to increase linearly with the fractional path length
$l(r)$ and express $DP_{\mathrm{screen}}(r)$ by:
\begin{equation}
DP_{\mathrm{screen}}(r) = l(r)\cdot (1-DP_{\mathrm{0}}) + DP_{\mathrm{0}}
\end{equation}
with $DP_{\mathrm{0}}$ the maximum intrinsic depolarization at $r=0$.
In this notation $DP=1$ means no depolarization and $DP=0$ complete
depolarization of the background radiation.

A Faraday screen also rotates the position angle of linearly polarized
background radiation. The amount of Faraday rotation is given by
$\Delta PA = RM\cdot\lambda^2$ with the rotation measure
$RM\,(\mbox{rad\, m}^{-2}) = 0.81\, B_{\parallel}\,
(\mu\mbox{G})\, n_{\mathrm{e}}\, (\mbox{cm}^{-3})\, l\,(\mbox{pc})$.
The rotation measure of the screen depends on the  fractional path
length $l(r)$ and can be expressed as follows:
\begin{equation}
RM_{\mathrm{screen}}(r) = RM_{\mathrm{0}}\cdot l(r)
\end{equation}
with $RM_{\mathrm{0}}$ the maximum intrinsic rotation measure at $r=0$.

Depolarization and Faraday rotation are the two effects a Faraday
screen might cause. The background polarized emission gets modified in
a systematic way as a function of $r$ and can be expressed by:
\begin{equation}
\begin{array}{rl}
PI_{\mathrm{mod}}(r)&=~DP_{\mathrm{screen}}(r) \cdot PI_{\mathrm{back}} \\
PA_{\mathrm{mod}}(r)&=~RM_{\mathrm{screen}}(r)\cdot\lambda^2 + PA_{\mathrm{back}}
\end{array}
\end{equation}

The observed polarization is the superposition of the modified
background and the foreground polarization, which in that case is a
vector rather than a scalar addition as in the case of total
intensities. This means to add $U$ and $Q$ and calculate from these
values the resulting $PI$ and $PA$. With
\begin{equation}
\begin{array}{rl}
U_{\mathrm{mod}}(r)&=~PI_{\mathrm{mod}}(r)\cdot\sin(2 PA_{\mathrm{mod}}(r))\\
Q_{\mathrm{mod}}(r)&=~PI_{\mathrm{mod}}(r)\cdot\cos(2 PA_{\mathrm{mod}}(r))
\end{array}
\end{equation}
The observable polarization calculates then by:
\begin{equation}
\begin{array}{rl}
PI_{\mathrm{obs}}(r)&=~\sqrt{\left(U_{\mathrm{fore}}+U_{\mathrm{mod}}(r)\right)^{2} +
\left(Q_{\mathrm{fore}}+Q_{\mathrm{mod}}(r)\right)^{2}}\\
PA_{\mathrm{obs}}(r)&=~\frac{1}{2}\arctan\left(\frac{U_{\mathrm{fore}}+U_{\mathrm{mod
}}(r)}{Q_{\mathrm{fore}}+Q_{\mathrm{mod}}(r)}\right).
\end{array}
\end{equation}

\begin{figure}[tph]
\centerline{\psfig{figure=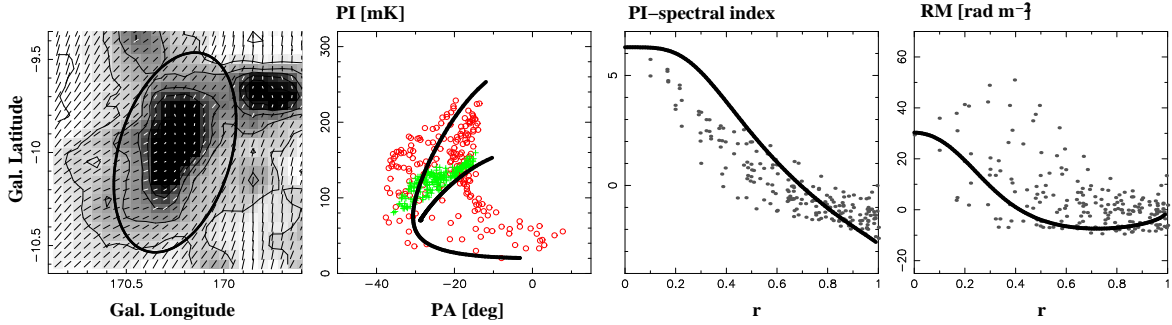,width=15.7cm,angle=0,clip=T,bbllx=10pt,bblly=244pt,bburx=470pt,bbury=377pt}}
\caption{From left to right are shown: map of the observed spectral
index of polarized emission with an ellipse marking the size of the
Faraday screen, the $PA$--$PI$ plot for $1408$~MHz (red) and $1713$~MHz
(green), the observed spectral index of polarized emission, as well as
the observed rotation measure versus radius $r$. The black lines
indicate the model-fit.}
\label{bildchenfit}
\end{figure}

Fitting the two modelled observables $PI_{\mathrm{obs}}$ and
$PA_{\mathrm{obs}}$ to the measured polarized intensities and angles
towards the Faraday screens is done by optimizing the four free
parameters of the model: $PI_{\mathrm{fore}}$, $PA_{\mathrm{fore}}$,
$RM_{\mathrm{0}}$, and $DP_{\mathrm{0}}$. At $r\ge1$ the pure
superposition of background and foreground polarization must result in
the measured polarization outside the screen. This limiting condition
constrains the background polarization for $r<1$. The model correctly
reproduces the observed high spectral indices of polarized intensities,
the rotation measures, as well as the observed variation in polarized
intensity and angle (see Fig.~\ref{bildchenfit}). We find the
following best-fit parameters for the Faraday screen marked in
Fig.~\ref{mapchen}: $PI_{\mathrm{fore}} \sim 150$~mK,
$PA_{\mathrm{fore}} \sim -1$\degs, $PI_{\mathrm{back}} \sim 130$~mK,
$PA_{\mathrm{back}} \sim -14$\degs, $RM_{\mathrm{0}} \sim
-29.3$~rad$\,$m$^{-2}$, and $DP_{\mathrm{0}}=1$. These values describe
the foreground and background emission, as well as the rotation
measure (see next paragraph) and depolarization at $1.4$~GHz. The
other Faraday screens which can be identified will be discussed in a
subsequent paper (Wolleben \& Reich 2004).

Limitations of this model arise from the simplification of the shape
and properties of Faraday screens, which are likely not perfectly
elliptical with constant electron density and homogeneous magnetic
field inside, but might be more turbulent or have a small filling
factor. However, the absence of depolarization indicates little
turbulence within the Faraday screen. Another simplification is that
shape and size of the screens were estimated by eyeball based on their
appearance in the PI--spectral index map. Finally, Faraday screens
can overlap, which was not accounted for, but which is probably the
case for the screen fitted here as seen from Fig.~\ref{mapchen}.

\section{Conclusions}

Since the Faraday screen we observed towards the Taurus--Auriga
molecular cloud complex is most likely associated with the molecular
gas, we can specify its distance to $140$~pc. When spherical symmetry
is assumed, the size of the screen is of the order of $2$~pc. An
intrinsic $RM$ of about $-29$~rad$\,$m$^{-2}$ requires an excessive
value for the thermal electron density or an excessive regular
magnetic field component parallel to the line-of-sight, when compared
to average Galactic values. The total power maps show no enhanced
thermal emission at $1408$, $1660$, or $1713$~MHz towards the Faraday
screen which gives an upper limit  for the thermal electron density of
$n_{\mathrm{e}}\le2$~cm$^{-3}$. In addition, we have checked available
H$\alpha$ data from the {\it full-sky H-alpha map} constructed by
Finkbeiner (2003). No enhanced emission (at the $1\sigma$ detection
level of $0.52$~Rayleigh) related to the molecular gas or the Faraday
screens is visible, which constrains $n_{\mathrm{e}}$ to less than
$0.8$~cm$^{-3}$ for electrons from hydrogen ionization. With these
upper limits for the thermal electron density a regular magnetic field
strength exceeding $20~\mu$G along the line-of-sight is needed to
explain the intrinsic $RM$.

Towards the Faraday screen modelled here, the observed rotation
measure $RM_{\mathrm{obs}}$  differs from its intrinsic rotation
measure $RM_{\mathrm{int}}$ by about $60$~rad$\,$m$^{-2}$ and the sign
of the observed $RM$s is in opposite direction. In the presence of
foreground polarization, which adds to the Faraday rotated background,
the observed $RM$ is not a fixed value, but depends on the two
frequencies used for its determination and in addition on the amount of
foreground polarization adding to the rotated background. This implies
no $\lambda^{2}$--dependence of the observed polarization angles in the
direction of Faraday screens.

\section*{References}\noindent

\references

Brouw W.~N., Spoelstra T.~A.~Th. \Journal{1976}{\AAps}{26}{129}.

Dame T.~M., Hartmann D., Thaddeus P. \Journal{2001}{\ApJ}{547}{792}.

Elias J.~H. \Journal{1978}{\ApJ}{224}{857}.

Finkbeiner D. P. \Journal{2003}{\ApJSS}{146}{407}.

Gom\'ez G.~C., Benjamin R.~A., Cox, D.~P. \Journal{2001}{\AJ}{122}{908}.

Reich P., Reich W. \Journal{1986}{\AAps}{63}{205}.

Reich P., Reich W. \Journal{1988}{\AAp}{74}{7}.

Reich W. \Journal{1982}{\AAps}{48}{219}.

Spoelstra T.~T.~Th. \Journal{1972}{\AAp}{21}{61}.

Taylor J.~H., Cordes J.~M. \Journal{1993}{\ApJ}{411}{674}.

Wolleben M., Reich W. (2004) {\em Astron. Astrophys.}, submitted.

\end{document}